# Deep Learning based Model-free Robust Load Restoration to Enhance Bulk System Resilience with Wind Power Penetration

Jin Zhao, *Member*, *IEEE*, Fangxing Li, *Fellow*, *IEEE*, Xi Chen, *Senior Member*, *IEEE*, Qiuwei Wu, *Senior Member*, *IEEE*

*Abstract*—This paper proposes a new deep learning (DL) based model-free robust method for bulk system on-line load restoration with high penetration of wind power. Inspired by the iterative calculation of the two-stage robust load restoration model, the deep neural network (DNN) and deep convolutional neural network (CNN) are respectively designed to find the worst-case system condition of a load pickup decision and evaluate the corresponding security. In order to find the optimal result within a limited number of checks, a load pickup checklist generation (LPCG) algorithm is developed to ensure the optimality. Then, the fast robust load restoration strategy acquisition is achieved based on the designed one-line strategy generation (OSG) algorithm. The proposed method finds the optimal result in a model-free way, holds the robustness to handle uncertainties, and provides real-time computation. It can completely replace conventional robust optimization and supports on-line robust load restoration which better satisfies the changeable restoration process. The effectiveness of the proposed method is validated using the IEEE 30-bus system and the IEEE 118-bus system, showing high computational efficiency and considerable accuracy.

*Index Terms*—Convolutional neural network (CNN), deep learning (DL), power system resilience, load restoration, wind power integration.

## NOMENCLATURE

### A. Indices

| | |
|---|---|
| $i$ | Index of nodes |
| $L_{inner}$ | Indices of iterative calculations |
| * | Index for variables with fixed values |
| $s$ | Index of samples |
| $k$ | Index of neurons |
| $j$ | Index of sample features |
| $l$ | Index of layers of neural networks |

### B. Sets

| | |
|---|---|
| $N_L$ | Set of load nodes |
| $N_W$ | Set of wind farm nodes |
| $N_s$ | Set of sample data |

### C. Parameters

| | |
|---|---|
| $\Delta f_{max}$ | Upper bound of frequency deviation |
| $V_{max}$, $V_{min}$ | Maximum and minimum values of voltage |
| $S_{max}$, $S_{min}$ | Set of sample data |
| $p_{low,i}$, $p_{up,i}$ | Lower and upper bounds of active power of node $i$ |
| $\mathbf{C}, \mathbf{D}, \mathbf{E}, \mathbf{F}$ | Matrixes of parameters of equalities |
| $\mathbf{H}, \mathbf{I}$ | Matrixes of parameters of in-equalities |
| $w_j^l$, $b_k^l$ | $j$th weight and $k$th biases parameters of $l$th layer |
| $\eta$ | Learning rate |
| $N_{Lay}$ | Total number of hidden layers |
| $\mathbf{P}, \mathbf{Q}, \mathbf{B}, \boldsymbol{\theta}, \mathbf{V}$ | Vectors of bus active and reactive power injection, susceptance matrix, voltage magnitude and angel |
| $L^{upp}$, $L^{pick,upp}$, $L^{pick,low}$ | Upper bound of the total pickup load amount, acceptable upper and lower bounds of load pickup amount |
| $s_j$, $\varepsilon_j$ | Capacity and frequency response rate of generator $j$ |
| $E_{W,i}$, $E_{W,i,ini}$ | Expected wind power output of wind node $i$ in the current step and last step |
| $I_{str}$ | The number of combinations in the checklist |

### D. Variables

| | |
|---|---|
| $\mathbf{X}_L$ | Vector of load pickup decisions |
| $\mathbf{E}_L$ | Vector of expected load amount |
| $\mathbf{p}_G$ | Vector of generator outputs |
| $\mathbf{p}_L$, $\mathbf{p}_W$ | Vectors of uncertain load amount and wind power output |
| $\mathbf{V}, \mathbf{S}$ | Vectors of voltage and branch power |
| $t$ | Restoration step time |
| $\boldsymbol{\lambda}, \boldsymbol{\mu}$ | Vectors of Kuhn–Tucker multipliers |
| $z_i^+$, $z_i^-$ | Auxiliary binary variables of node $i$ |
| $A_I^+, A_I^-, B_I^+, B_I^-$ | Auxiliary variables and binary variables for linearization |
| $x_j^l$ | $j$th input of $l$th layer |
| $z_k^l$, $y_k^l$ | $k$th weighted aggregation and output of $l$th layer |
| $p_{L,i}^s$, $p_{W,i}^s$, $p_{G,i}^s$, $q_{G,i}^s$ | Variables of load and wind active power and generator active and reactive power for $s$th sample of node $i$ |
| $V_i^s$, $\theta_i^s$ | Variables of voltage magnitude and angel for $s$th sample of node $i$ |

J. Zhao and F. Li are with the Department of EECS, The University of Tennessee, Knoxville. TN 37996, USA (e-mail: jzhao44@utk.edu; fli6@utk.edu).
X. Chen (e-mail: xi.chen@geirina.net).
Q. Wu (e-mail: quiwudtu@gmail.com).

## I. Introduction

HIGH-IMPACT and low-probability events, such as extreme weather events, are occurring with increasing intensity [1], indicating the necessity of enhancing power system resilience. After a complete or partial collapse of a power system, power system restoration is crucial to bringing the system back to normal conditions [2], [3]. Fast load restoration, as the final goal of the restoration process, is indispensable to power system resilience after outage events [4]. In this process, the resilience is defined as the ability to quickly pick up shed loads and is measured by recovered load amount. At the bulk system level, the essential purpose of large-scale load restoration is to efficiently and safely dispatch available power sources to satisfy the power demand of load blocks. Since the generation is limited in the restoration process [5], the optimal load restoration strategy is pursued to make full use of the available power supply in the system. In a step-by-step process, there are three major issues for determining the load restoration strategy of each step: i) location of load to restore, i.e. where, ii) restorable size of load, i.e. how much, and iii) operation time, i.e. when [6]. Mathematically, it is a combination optimization problem. With binary variables representing load pickup decisions, the mixed-integer optimization [7], [8] is widely used to find the optimal load restoration strategy.

With increasing integration in power systems, renewable energy sources (RESs) have become an indispensable part of the power supply in the restoration process. Wind power based sources benefit load recovery with fast self-restarts and a flexible power supply [9]. However, because of the intermittency of wind power sources and the cold pickup phenomenon, uncertainties should be considered. Therefore, load restoration optimization in a wind power penetrated system requires uncertainty handling capabilities to guarantee the feasibility of the obtained scheme [10]-[13]. In [10], stochastic programming was used to develop an offline restoration planning tool for harnessing wind energy to enhance grid resilience. The chance-constrained stochastic program was applied in [11] to model the load restoration problem considering stochastic energy variations. The robust optimization method helps realize restoration related coordination in a wind power participated restoration process [12], [13].

In order to achieve optimal restoration decision-making, model-based methods [7]-[13] have been well studied in recent works. However, model-based optimization methods are generally time-consuming in dealing with uncertainties. The computation efficiency of the stochastic programming is low when the number of scenarios is large [10], [11], and robust restoration models require iterative calculation which increases model scale in the iteration [12], [13]. The time-consuming model solving limits the corresponding strategies to off-line applications [10], [13], which significantly reduces their practicality. Therefore, fast acquisition of a robust load restoration strategy is needed to realize on-line robust restoration with real-time computation.

Deep learning (DL) methods, such as deep neural networks (DNNs) and deep convolutional neural networks (CNNs), promise to realize on-line robust load restoration. In a model-free way, DL methods automatically extract features from provided data sets and achieve accurate model regression [14]. Well-trained DNN and CNN develop high generalization and can be directly applied to new cases without costly numerical computation. Accordingly, in recent years, DL methods have been applied to help solve various problems in power systems [15]-[17]. DNN has been utilized in load and wind power forecasting [18], RES modeling [19] and optimization model regression [20], and the efficiency of deep CNNs has been verified in sparse vector related classification [21] and AC power flow (PF) equations related regression [22]. In the field of power system restoration, artificial neural networks (ANNs) were once popular [23], however, the application of DL methods is still in the initial stage. With high computational efficiency and considerable accuracy, the DL based method has great potential to satisfy the real-time computational requirements of on-line robust load restoration.

Existing model-based methods with uncertainty considered are generally limited to off-line applications, while on-line load restoration is preferred in the wind power penetrated restoration process. Therefore, a new deep learning-based model-free robust load restoration method is proposed in this paper to support the real-time decision-making of the on-line load restoration process. First, the two-stage robust optimization of load restoration is performed to provide training data for the DL process. Then, a DNN and CNN are designed and trained to capture the worst-case condition and check the security of load pickup decisions under uncertainties. Finally, a load pickup checklist generation algorithm (LPCG) and a one-line strategy generation (OSG) algorithm are developed to find the final optimal robust load restoration strategy in a model-free way. The whole methodology contains different approaches, although the model-based method is applied to complete the training of DL methods, the proposed method finally realizes the model-free load restoration process.

The contributions of this paper can be summarized as follows: 1) This work applies advanced DL methods in the field of bulk power system restoration. Differing from the conventional model-based methods, this paper proposes a new model-free method for finding the optimal load restoration strategy. The well-studied optimization techniques provide training data for the DL method, and well-trained neural networks are used in the load restoration process to realize the model-free load restoration strategy generation. 2) The proposed method is with uncertainty handling ability. The original problem is decomposed into three sub-problems which fit into model-free solutions. Then, a series of techniques, including a DNN, a CNN, the LPCG algorithm and the OSG algorithm, are designed and combined to fully replace the robust mathematical optimization process. With good optimization performance, the obtained result is robust to uncertain load pickup amount and wind power output, which ensures safe restoration of the wind power penetrated system. 3) Finally, this work realizes the on-line application of the robust load restoration scheme. With high computational efficiency and considerable accuracy, the proposed method provides the real-time computation of the robust strategy. This allows the proposed method to apply at the on-line level to better satisfy the changeable restoration process.

The rest of the paper is organized as follows: Section II provides the two-stage robust load restoration models and the corresponding solution method. The DNN and CNN are designed and trained in Section III. Section IV shows the developed LPCG algorithm and OSG algorithm. Section V provides case study results and discussions, followed by the conclusions.

## II. Optimization models of robust load restoration

This section introduces the robust optimization of the load restoration problem. The two-stage robust optimization model is constructed followed by a column and constraint generation (C&CG) based iterative calculation process.

### A. Robust load restoration model

The goal of load restoration is to quickly restore sufficient power supply to all load blocks via a step-by-step process. The objective of each step is maximizing the load recovery amount in a meshed network considering the limited generation of generators, line power flow limits, and voltage and frequency securities. In the load restoration process, the cold load pickup (CLPU) phenomenon caused by thermostatically controlled loads leads to unpredictable inrush load pickup amount. Due to uncertainties of wind power output and load amount caused by the cold load pickup phenomenon, the robustness of the load restoration strategy is enhanced using the two-stage robust form (1). Model (1) is a single-step optimization model which is recursively performed in the step-by-step load restoration process.

$$\min_{X_L} -sum(X_L E_L^T) + \max_{p_L, p_W} \min_{p_G, t, V, S} \left( sum(p_G) + sum(p_W) \right) \quad (1a)$$

$$s.t. \quad F_1(X_L) \leq 0 \quad (1b)$$

$$G(p_G) = F_2(t) \leq 0 \quad (1c)$$

$$\Delta f = F_3(p_L, p_W) \leq \Delta f_{max} \quad (1d)$$

$$\begin{cases} [V, S] = F_4(p_G, p_L, p_W) \\ V_{min} \leq V \leq V_{max} \\ S_{min} \leq S \leq S_{max} \end{cases} \quad (1e)$$

The objective (1a) contains two terms. The first term of (1a) is the weighted expected load restoration amount and the latter term represents the power supply requirement of the restored load. In the *max-min* term, the *max* finds the worst case condition w.r.t uncertain variables $p_L$, $p_W$, while *min* denotes reducing the power supply requirement under the worst case w.r.t $p_G$, $t$, $V$, and $S$. In load pickup constraint (1b), $F_1(.)$ represents decision variable related functions which describe the practice load pickup characteristic. $F_2(.)$ and $G$ () provide the function relationship between step time and generator outputs. Accordingly, constraint (1c) is used to satisfy generator ramping up characteristics [5] as well as the spinning reserve which survives the largest loss of energy contingency [24]. The frequency deviation and dynamic reserve limits are presented in (1d) which contains $F_3(.)$ reflecting the relationship between the system active power increment and frequency deviation [25]. Voltage and branch flow limits are included in the AC PF related constraints (1e) where $F_4(.)$ denotes power flow functions.

Normally, power flow calculation will be convexified to realize tractable computation of load restoration optimization. Without loss of generality, this paper takes the linearized AC PF calculation method as an example [26]. Other convexification methods can also be used, and the feasibility of the proposed method will not be influenced.

### B. Solution of model-based robust load restoration

Model (2), which is a two-stage robust optimization with a polyhedral uncertainty set $\Phi$ (2d) for uncertain wind power output and load amount [27], is the equivalently transformation of model (1). Therein, constraint (1b) is represented by (2b) and (1c) – (1e) are included in (2c).

$$\min_{X_L} -X_L E_L^{'T} + \max_{p_L, p_W} \min_{Y} \left( sum(p_G) + sum(p_W) \right) \quad (2a)$$

$$s.t \quad \Delta = \{X_L : (1b)\} \quad (2b)$$

$$\Psi = \{Y = (p_G, V, S) : (1c), (1d), (1e)\} \quad (2c)$$

$$\Phi = \{(p_L, p_W) : p_{low,i} \leq p_i \leq p_{up,i}, i \in N_L \& N_W\} \quad (2d)$$

The first stage includes the first *min* term of the objective in (2a) and load pickup variable related constraints (2b), while the second stage contains the *max-min* term in the objective (2a) and all the remaining constraints (2c)-(2d). The robust load restoration maximizes load recovery in the first stage and minimizes the generation requirement in the worst-case condition under uncertainties. In order to solve the integer variable integrated two-stage problem, the C&CG method is applied with an iterative calculation of inner and outer sub-models.

According to the C&CG method, the inner-level sub-model (2) is shown as model (3). In (3), (3b) and (3c) denote equality and inequality of the second stage constraint (2c). Model (3) is further transformed into model (4) using dual theory for the purpose of solution. Note that the dual theory is used when the AC PF is linearized. In other convexification cases, the Karush–Kuhn–Tucker (KKT) condition can be used to get model (4).

$$\max_{(P_L, P_W) \in \Phi} \min_{Y} \left( sum(p_G) + sum(p_W) \right) \quad (3a)$$

$$s.t. \quad CY + DX_L^* p_L + E p_W + F = 0 \ : \mu \quad (3b)$$

$$HY + I \leq 0 \ : \lambda \quad (3c)$$

$$\max_{(P_L, P_W) \in \Phi} DX_L^* p_L \mu + E p_W \mu + F \mu + I \lambda \quad (4a)$$

$$s.t. \quad \mu C + \lambda H = 0 \quad (4b)$$

$$\lambda \geq 0 \quad (4c)$$

With determined load pickup decision $X_L^*$ obtained from the outer-level problem, the goal of the inner-level problem is providing the worst-case condition for the outer-level problem. Since it is proven that the worst-case realization of the polyhedral uncertainty is identified at vertices, (5) holds in model (4).

$$\begin{aligned} p_i \mu_i &= \left( z_i^+ p_{up,i} + z_i^- p_{low,i} \right) \mu_i \\ s.t. \ z_i^+ + z_i^- &= 1, \forall i \in N_L \& N_W \end{aligned} \quad (5)$$

Using the Big M method, (5) is linearized as (6).

$$p_i \mu_i = A_i^+ p_{up,i} + A_i^- p_{low,i} \quad (6a)$$

$$s.t. \begin{cases} -B_i^+ M \leq A_i^+ \leq B_i^+ M \\ -(1-B_i^+)M \leq A_i^+ - \mu_i \leq (1-B_i^+)M \\ -B_i^- M \leq A_i^- \leq B_i^- M \\ -(1-B_i^-)M \leq A_i^- - \mu_i \leq (1-B_i^-)M \\ B_i^+ + B_i^- = 1, \forall i \in N_L \& N_W \end{cases} \quad (6b)$$

M is a big enough constant, $A_i^+$ and $A_i^-$ are auxiliary variables to linearize (5), and $B_i^+$ and $B_i^-$ are the corresponding binary variables.

Accordingly, the inner-level problem is finally transformed into the solvable mixed-integer linear programming (MILP) problem from (7).

$$\max \sum_{i \in N_L} \left[ D_i x_{L,i}^* \left( A_i^+ p_{up,i} + A_i^- p_{low,i} \right) \right] \\ + \sum_{i \in N_W} \left( A_i^+ p_{up,i} + A_i^- p_{low,i} \right) + \mathbf{F}\boldsymbol{\mu} + \mathbf{I}\boldsymbol{\lambda} \quad (7)$$

$$s.t. \ (3b), (3c), (6b)$$

Adding new extra variables in $\mathbf{Y}^l$ and the corresponding constraints in (8c), the outer-level sub-model is shown in (8) with the fixed worst-case $\boldsymbol{p}_L^*$ and $\boldsymbol{p}_W^*$ provided by the solution of the inner-level problem. Therein, $L_{inner}$ represents the number of iterative calculations.

$$\min_{x_L, \vartheta, Y^l} -X_L E_L^{\prime T} + \vartheta \quad (8a)$$

$$s.t. \ X_L \in \Delta \quad (8b)$$

$$\begin{cases} \vartheta \geq sum(\boldsymbol{p}_G^l) + sum(\boldsymbol{p}_W^*) \\ \mathbf{C}Y^l + \mathbf{D}X_L \boldsymbol{p}_L^* + \mathbf{E}\boldsymbol{p}_W^* + \mathbf{F} = 0 \\ \mathbf{H}Y^l + \mathbf{I} \leq 0 \\ \forall l \in \{1,2...L_{inner}\} \end{cases} \quad (8c)$$

Model (8) provides the determined load pickup decisions $X_L^*$ for model (7), and model (7) feeds back the worst-case condition for model (8). The final optimal robust result is obtained as long as the iterative calculations of (7) and (8) converge.

It should be noted that the essence of solving the two-stage robust optimization is to find the worst-case condition of a selected load pickup decision. Then, based on the corresponding constraints, the load pickup decision under the worst-case condition is updated. The proposed model-free method is inspired by this process. Optimization models in this section provide training data for neural networks used in the DL method that will be elaborated next.

## III. DL BASED UNCERTAINTY HANDLING METHOD

In this section, a DNN and CNN are designed and trained to replace the model optimization process. The DNN is used to determine the worst-case condition of a load pickup decision under uncertainties, while the CNN checks the corresponding security of the worst-case condition.

### A. Find worst-case condition using DNN

According to the *Universal Approximation Theorem*, the neural network has the ability to approximate any continuous functions [28]. For the optimization model, there is actually a complex function relationship between the input and output data. Therefore, theoretically, a DNN can mimic convex model optimization with considerable accuracy.

As shown in Fig. 1, the input data to the DNN is the expected values of wind power output and the expected load pickup amount. The output data includes the worst-case condition of uncertain load and wind variables ($\boldsymbol{p}_L^*$ and $\boldsymbol{p}_W^*$) and the corresponding generator output condition. The goal of the DNN is to generate the worst-case condition of a load pickup decision under uncertainties. The result will be used for security checks.

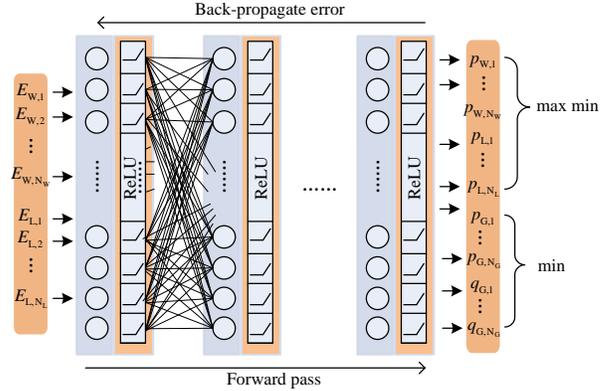

Fig. 1 Structure of DNN for robust load restoration

1) DNN data processing

The training data set for the DNN comes from two optimization models in Section III. The first one is the inner-level sub-model (7) and the second one is the original load restoration model (1) with fixed $X_L^*$, $\boldsymbol{p}_L^*$ and $\boldsymbol{p}_W^*$. The expected values of wind power output $E_{W,i}$ are obtained from system data. With a selected load pickup decision $X_L^*$, the expected values of load $E_{L,i}$ can be determined. The selected load pickup decision $X_L^*$ is the one in the load pickup checklist (LPC) which will be in introduced later in Section V. A. The worst load amount $\boldsymbol{p}_L^*$ and wind power output $\boldsymbol{p}_W^*$ are the results of solving (7). Using fixed $X_L^*$, $\boldsymbol{p}_L^*$ and $\boldsymbol{p}_W^*$, model (1) becomes a linear programming (LP) form and provides generator output values $p_G$ and $q_G$.

In order to improve computational efficiency, a *min_max_scaler* transformation in (9) is applied to normalize the input and output data. Through the normalization, the values of the data are within the range of [0,1], which helps create a more regular search region for the faster convergence of the algorithm. In (9), D means the training data.

$$D_{i,s} = \frac{(D_{i,s}^0 - \min \boldsymbol{D})}{(\max \boldsymbol{D} - \min \boldsymbol{D})} (i \in N_L \text{ or } N_W, s \in N_S) \quad (9)$$

In the DNN structure of Fig. 1, the neurons realize the affine transformation of the input (10), and active functions (11) bring nonlinearity.

$$z_k^l = \sum_{j \in J} \left( \omega_j^l x_j^l + b_k^l \right) \quad (10)$$

$$y_k^l = f\left(z_k^l\right) \quad (11)$$

where $z$ is the weighted aggregation of all the features of the input *x* captured by neurons, *y* is final output of a layer, and *w* and *b* are weights and biases parameters. In this study, the rectifier linear units (ReLU) function is used as the activation function. The ReLU function has the form: $f(x) = \max(x,0)$ which preserves strong generalization abilities and allows easy application of gradient-based methods [14].

2) Training DNN

Getting well trained DNN is to find fine-tuned network parameters *w*, *b* using the back-propagation algorithm. The loss function minimizes the deviation between the data obtained from optimization models and the estimated value using the DNN. In this way, an accurate enough approximation of the optimization models can be achieved as long as the loss function value is small enough.

The loss function (12), which implies the accuracy of the output of the DNN, is defined as the objective of the DNN training. Because the outputs are all continuous values, the mean squared error (MSE) loss function is applied:

$$L_{DNN}(\omega_{DNN}, b_{DNN})$$
$$= \frac{1}{N_S} \sum_{s \in N_S} \left[ \frac{1}{N_W} \sum_{i \in N_W} \left( p_{W,i}^{s*} - p_{W,i}^{s} \right)^2 + \frac{1}{N_L} \sum_{i \in N_L} \left( p_{L,i,s}^{s*} - p_{L,i}^{s} \right)^2 \right. $$
$$\left. + \frac{1}{N_G} \sum_{i \in N_G} \left( p_{G,i}^{s*} - p_{G,i}^{s} \right)^2 + \frac{1}{N_G} \sum_{i \in N_G} \left( q_{G,i}^{s*} - q_{G,i}^{s} \right)^2 \right] \quad (12)$$

where the sample data $p_L^{s*}$ and $p_W^{s*}$ and $p_G^{s*}$ and $q_G^{s*}$ are respectively obtained from MILP model (7) and model (3) in the LP form as explained in the previous subsection, 1) DNN data processing.

Accordingly, the first partial derivatives of the loss function (12) with respect to *w* and *b* are used. Following the chain rule, the gradient of updating weights and biases is found by (13). Then, the final result can be obtained through the gradient descent search based iteration.

$$\omega_{i+1}^l = \omega_i^l - \eta \frac{\partial L}{\partial y_i^{N_{Lay}-1}} \cdot \frac{\partial y_i^{N_{Lay}-1}}{\partial y_i^{N_{Lay}-2}} \cdot \ldots \cdot \frac{\partial y_i^l}{\partial \omega_i^l} \quad (13)$$

*i* denotes the iteration number and $y_i^l$ is the output of DNN layer *l*. The bias has the same updating form.

With (13), the back-propagation algorithm uses the gradient to guide the updating of the neural network parameters toward the global optimum. Consequently, the well trained DNN can be built when the weights and biases are determined.

3) Accuracy improvement

In Fig.1, the first part of output replaces the result of *max-min* optimization. According to the *max-min* optimization theory, the worst-case realization of polyhedral uncertainty is identified at the vertices. Therefore, this part of result can be further corrected to reach the nearby vertices. In other words, the DNN finds the result near the exact optimal point, and then the output data can be further moved to the optimal point using the optimization theory. In this way, the worst-case load amount $p_L^*$ and wind power output $p_W^*$ obtained by the DNN can easily reach 100% accuracy.

B. *Security check of worst-case condition using CNN*

The deep CNN is known for its strong automatic feature learning ability in processing data with a grid-like topology, e.g., image data. According to the AC PF calculation (14), state variables such as bus voltage magnitudes and bus voltage angles, are mainly related to neighboring variables and parameters. This characteristic makes power system state variables sparsely connected, similar to image pixels, which is especially suitable for the application of the deep CNN [22]. Accordingly, the CNN can provide AC PF results that can be used to check voltage and branch power flow securities.

1) CNN structure

The structure of the deep CNN for voltage and angle calculation is illustrated in Fig. 2. It consists of two convolutional (Conv) layers and a fully-connected (FC) layer. The convolutional layers use function (15) to conduct convolution operations for feature extractions. The ReLU active functions (16) are used to add non-linearity.

$$\begin{cases} P_{nm} = V_n^2 g_{nm} - V_n V_m \left( g_{nm} \cos\theta_{nm} + b_{nm} \sin\theta_{nm} \right) \\ Q_{nm} = -V_n^2 b_{nm} - V_n V_m \left( g_{nm} \sin\theta_{nm} - b_{nm} \cos\theta_{nm} \right) \end{cases} \quad (14)$$
$$\forall n, m \in N, n \neq m$$

$$z_{i,j} = \sum_{u=1}^{m} \sum_{v=1}^{n} \left( \omega_{u,v} x_{i+u-1, j+v-1} \right) + b \quad (15)$$

$$y_{i,j} = f(z_{i,j}) \quad (16)$$

where $x_{i+u-1, j+v-1}$, $z_{i+v-1, j}$ and $w_{u,v}$ are a single unit (a single square in Fig. 2) of the input data, output data and convolution kernel square. *m* and *n* are the height and width of a filter. All the output units constitute a feature map which contains the extracted information using the $m \times n$ filter. With more filters, more feature maps can be generated to provide sufficient information for deep CNN model regression.

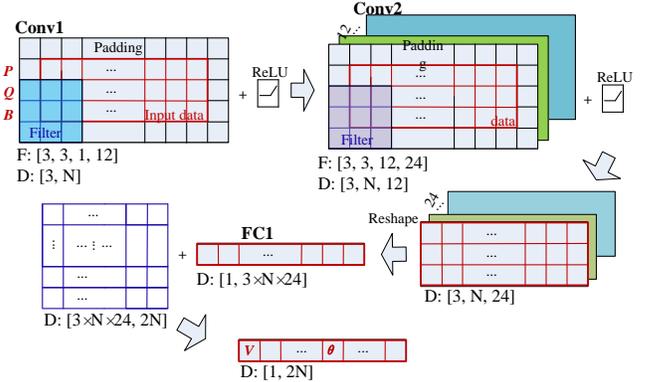

Fig. 2 Structure of CNN for power flow calculation

2) CNN data processing

The input data is obtained by the DNN in the last subsection, and it contains bus active power injection vector **P**, bus reactive power injection **Q** and self-susceptance elements of bus susceptance matrix **B**. The output data contains bus voltage magnitude vector **V** and angel vector ***θ***. Accordingly, the size of the input data [**P**; **Q**; **B**] for the deep CNN is $3 \times N$, and the size of the output data [***θ***; **V**] is $1 \times 2N$. The *min_max_scaler* transformation in (9) is applied to normalize the data. The technique of zero-padding is applied to maintain the original size of the input data.

3) Training CNN

The loss function (17) is defined as the objective of the CNN training. Therein, $V_i^{s*}$ and $\theta_i^{s*}$ are obtained by calculating (14) based on the system P-Q condition determined according to the DNN output data.

$$L_{CNN}(\omega_{CNN}, b_{CNN}) = \frac{1}{N_S} \cdot \frac{1}{N} \sum_{s \in N_S} \sum_{i \in N} \left[ (V_i^{s*} - V_i^s)^2 + (\theta_i^{s*} - \theta_i^s)^2 \right] \quad (17)$$

Similar to the DNN training process, the deep CNN is trained using the back-propagation method to find the fine-tuned network weight parameters and bias parameters.

In short, the DL based uncertainty handling approach consists of a DNN and a CNN. They are trained to mimic the model optimization and AC PF calculation. In a model-free way, the well-trained DNN can quickly determine the worst-case condition, and the well-trained CNN can quickly provide the AC PF result of the worst-case condition.

## IV. OPTIMAL ROBUST LOAD RESTORATION STRATEGY

Two algorithms are developed in this section. The LPCG algorithm is designed to generate the load pickup checklist which helps capture the optimality. The OSG algorithm is used to realize on-line optimal robust load restoration acquisition in a DL based model-free way.

### A. Load pickup checklist generation

The load pickup decision takes the form of 0/1 combinations representing the reconnection of load blocks. Therefore, a direct idea to find the optimal load pick up decision is to enumerate all the combinations and check their security constrains using the DL method described in Section III. The optimal robust decision is the one with the largest expected load pickup amount as well as no security violations. However, the number of combinations increases exponentially with the number of load blocks. For a large-scale system with numerous load blocks, it is problematic to list and check all the combinations. Therefore, the LPCG algorithm is developed to help find the optimal robust load restoration strategy within a limited number of checks.

In the load restoration process, frequency security is a relatively strict constraint for the optimization model. Normally, the optimal result falls near the boundary of the frequency constraint. Since the frequency calculation is related to system active power, it can be used to find the upper bound of the total load pickup amount. In this paper, the linearized frequency calculation method [10], [24] is used to gives the upper bound of the total pickup load amount $L^{upp}$ in (18). Other function relationships of load mount and frequency, such as [28], can also be effectively applied.

$$L^{upp} \leq \Delta f_{max} \sum_{j \in N_G, j \neq i} \frac{s_j}{\varepsilon_j} + \sum_{i \in N_W} (E_{W,i} - E_{W,i,ini}) (i \in N_G) \quad (18)$$

Using the simple optimization model (19), the load pickup combination with the smallest distance to the upper bound $L^{upp}$ can be found as $X_L^0$. This result is better than or equal to the optimal result, and it will be used as the first set of load pickup decisions in the LPC.

$$\min \left[ \sum_{i \in N_L} (E_{L,i} x_{L,i}) - L^{upp} \right]^2 \quad (19)$$

Holding $X_L^0$, an LPC with the decreasing load pickup amount is generated using the LPCG algorithm.

---

**LPCG algorithm:** Generate LPC

**Input:** The maximum load pickup amount $L^{upp}$, the acceptable lower bound $L^{pick,low}$ and the maximum number of combinations $I_{max}$.

**Output:** LPC

$S$1: Set the maximum load pickup amount $L^{upp}$ as $L^{pick,upp}$. Initialize LPC as $\{X_L^0\}$. Set $I_{str}$=0.

$S$2: **for** $I_{str}$ from 0 to $I_{max}$ **do**
    Set $L^{pick,upp} = X_L^{I_{str}} E_L$. Solve model (20)

$$\min \left[ \sum_{i \in N_L} (E_{L,i} x_{L,i}) - L^{pick,upp} \right]^2 \quad (20a)$$

$$s.t. \quad L^{pick,low} \leq \sum_{i \in N_L} (E_{L,i} x_{L,i}) \leq L^{pick,upp} - \sigma \quad (20b)$$

    and get the load pickup decision as $X_L^{I_{str}+1}$.
    **if** $X_L^{I_{str}+1}$ is equal to $X_L^{I_{str}}$.
      **break;**
    **else**
      Update the LPC as $\{X_L^0, X_L^1, ..., X_L^{I_{str}+1}\}$. Set $I_{str}$= $I_{str}$ +1.
    **end**
**end**

$S$3: Obtain the load pickup decision list $\{X_L^0, X_L^1, ..., X_L^{I_{str}}\}$.

---

Therein, σ in (20b) is a small enough constant (e.g. 1*10^(-4)). The finally obtained LPC { } stores load pickup decisions with decreasing expected load recovery amounts.

### B. On-line optimal robust strategy acquisition

By now, the security check of a load pickup decision under uncertainties can be realized by DL methods, and a load pickup checklist has been generated. Since the load pickup decisions in the LPC are ordered from largest load pickup amount to the lowest one, the one-by-one security check can be performed, and the first feasible set is the optimal one with the largest load pickup amount. Moreover, because the worst-case conditions of the decisions in LPC are considered using the DNN and CNN, the result is with the robustness to handle uncertainties. In order to apply the DNN, CNN and LPC in the restoration process, the OSG algorithm is designed to support the on-line load restoration with fast robust strategy generation.

The whole methodology of the proposed method is shown in Fig. 3. The training of the DNN and CNN and the LPCG algorithm are off-line pre-processing, while the application of the well-trained DNN and CNN and generated LPC is on-line. Without any optimization model being built and solved, the calculations in the OSG algorithm based on-line process can be extremely fast. Therefore, the optimal robust strategy acquisition can be a real-time application, and the DL technique enables the model-free on-line decision-making process to completely replace the two-stage robust optimization.

**OSG algorithm:** Find the optimal robust result
  **Input**: LPC and well trained DNN and CNN
  **Output**: The optimal robust strategy $[X_L^*, p_G^*]$.
  S1: Organize input data $E_L$ and $E_w$ for the well trained DNN, and get the worst-case condition containing $[p_L, p_w, p_G, q_G]$ for LPC.
  S2: Check frequency security based on (21) which follows same frequency calculation method of (18) and replaces expected values in (18) with corresponding worst-case values in $[p_L, p_w, p_G, q_G]$. Delete infeasible load pickup decisions in LPC and obtain an updated LPC2.

$$\Delta f = \left[ sum(p_L) - sum(p_w - p_{ini}) \right] \bigg/ \sum_{j \in N_G, j \neq i} \frac{s_j}{\varepsilon_j} \quad (i \in N_G) \quad (21)$$

  S3: Define system P-Q condition, and apply the well-trained CNN to get $[\theta, V]$ for LPC2.
  S4: **for** $I$ from 0 to $I_{max2}$ **do**
    Get $X_L^I$ from LPC2, and get the corresponding $p_G$, $[\theta^I, V^I]$ and $S^I$ of $X_L^I$.
    **if** voltage, branch flow and generator output limits are satisfied
      **break;**
    **else**
      $I = I + 1$.
    **end**
  S5: Obtain the optimal robust load restoration strategy from LPC2.

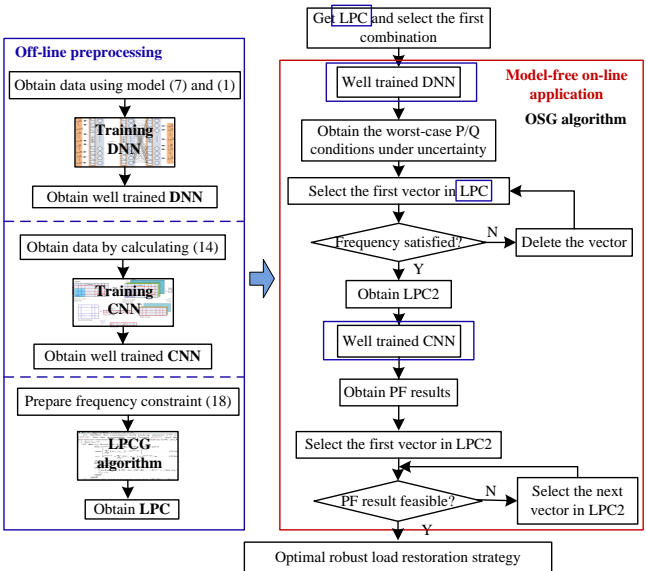

Fig. 3 Flowchart of the DL based robust load restoration method

## V. CASE STUDY

In this section, the testing performances of the DNN and CNN are presented. Then, the LPCG algorithm and OSG algorithm are applied to decide the optimal robust load restoration strategy. The results are evaluated and compared with the conventional model-based method to demonstrate the advantages of the model-free method.

Two modified systems are used: the five wind farms integrated IEEE 30-bus system and ten wind farms integrated IEEE 118-bus system. 10% and 20% deviations from the expected values are used to determine the bound of polyhedral uncertainty sets of load amount and wind power output, respectively. The optimization models were solved using CPLEX V12.5.1 on a computer with Intel(R) Core(TM) i7-8550U CPU and 16 GB RAM. The DNN and CNN are built and trained using TensorFlow 1.15 implemented on Python 3.7.

### A. Computation performance of DL method

#### 1) DNN training and test results

A DNN with 3 hidden layers is designed to find the worst-case condition. The neuron number is 1000, and the learning rate is 0.001. The size of the training data for the IEEE 30-bus system is 1500 while the size of the test data is 500. For the IEEE 118-bus system, the training data size and test data size are 8000 and 2000, respectively. The training/test data is generated using MILP model (7) and model (1) in LP form as explained in Section III *A* 1). For the IEEE 30-bus system, the number of inputs is 35 including expected load pickup amounts at 30 nodes and expected wind power output amounts at 5 nodes. The corresponding number of outputs is 55 including load pickup amount, wind power output and generator output in the worst-case condition. For the IEEE 118-bus system, numbers of inputs and outputs are 128 and 166. In the DNN, the input data represents a selected load pickup decision and the output data shows the worst-case P-Q condition in the system under uncertainties.

Table I shows the training result of the DNN for two systems. As shown in Fig. 4, after 200 iterations, the training loss value for the small-scale first system is 0.000026 which is small enough to indicate the training convergence. The final accuracies in training set and test set are respectively 99.45% and 99.38%, which verify the considerable accuracy of DNN regression.

TABLE I. TRAINING RESULTS OF DNN

| Case | Iteration | Loss value | Train | Test |
|---|---|---|---|---|
| IEEE-30 | 200 | 0.000026 | 99.45% | 99.38% |
| IEEE-118 | 500 | 0.000309 | 98.40% | 96.48% |

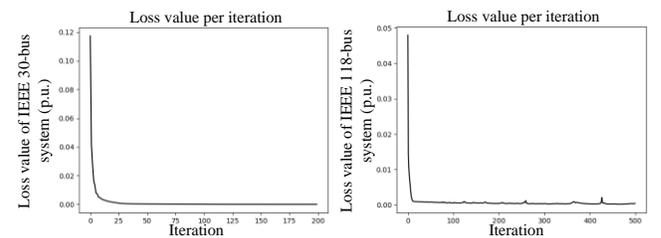

Fig. 4 Regression of the worst-case using DNN

Using the same DNN structure, the training loss value for the second system is 0.000309 after 500 iterations. The corresponding accuracies are 98.40% and 96.48% for training data and test data, respectively. If needed, the test accuracy can be further improved by adding $L_2$ regularization in the loss function. Here, because the worst-case condition will be corrected later, this accuracy is acceptable and it has limited effect on the final result.

*2) Worst-case condition correction*

The output of the DNN provides a result near the optimal point. The result can be further corrected to the strictly optimal points of the *max-min* robust model. According to the optimization principle, the worst-case realization is identified at the vertices for polyhedral uncertainty. Taking five DNN examples for example, Fig. 5 shows the correction process. The DNN results (blue points) are further moved to nearby vertices (red points). Through the correction, the accuracy of worst-case load amount and wind power output can reach 100%, which benefits the later security check.

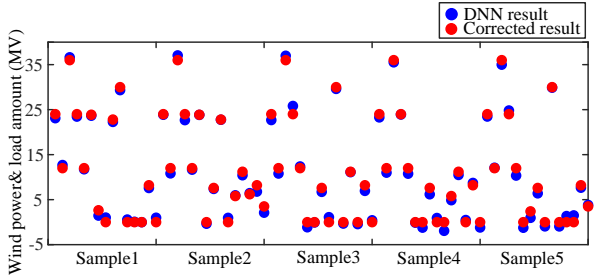

Fig. 5. Further correction of DNN result

*3) CNN training and test results*

The CNN with two convolutional (Conv) layers and one fully-connected (FC) layer is designed to realize an AC PF based security check. The filter of Conv1 has the size [3, 3, 1, 12] and the one of Conv2 has the size [3, 3, 12, 24]. The FC layer has 3000 neurons. The learning rate is 0.001. Numbers of input and output data are $3 \times 33$ and 66 for the first system and $3 \times 118$ and 236 for the second system. The input data is the susceptance matrix **B** and the worst-case P-Q system condition obtained according to the output data of DNN. The output is the voltage and angle of each node in the system generated using MATPOWER.

TABLE II. TRAINING RESULTS OF CNN

| Case | Train size/test size | Loss value | Train_$V$ | Test_$V$ | Train_$\theta$ | Test_$\theta$ |
|---|---|---|---|---|---|---|
| IEEE-30 | 1500/500 | 0.000134 | 99.82% | 99.79% | 99.82% | 98.92% |
| IEEE-118 | 8000/2000 | 0.000199 | 99.82% | 99.82% | 99.82% | 99.65% |

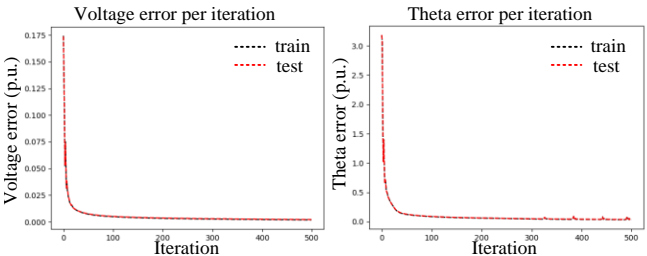

Fig. 6. Regression of AC PF of IEEE 30-bus system using CNN

Table II shows the training result of the CNN. Iteration processes for the two systems are shown in Fig. 6 and Fig. 7. After 500 iterations, the accuracies of obtained voltage and angle are around 99% which satisfies the security check.

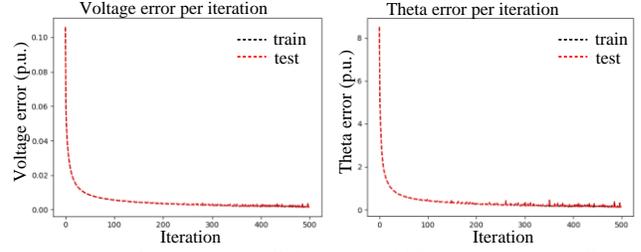

Fig. 7. Regression of AC PF of IEEE 118-bus system using CNN

*B. Load restoration performance of model-free method*

The off-line training of the DL based model-free method can be completed by the process above. Then, the LPCG algorithm is used to generate the LPC.

*1) LPC generation*

As shown in Fig. 8, the LPC gives several combinations of load pickup decisions from the largest amount to the lowest amount. The combinations become a LPC which will be checked one-by-one using the OSG algorithm. The computation time of generating sixty LPCs is 10.12s for the first system, and 1074.15s for the second system. It is acceptable because the LPC can be prepared off-line.

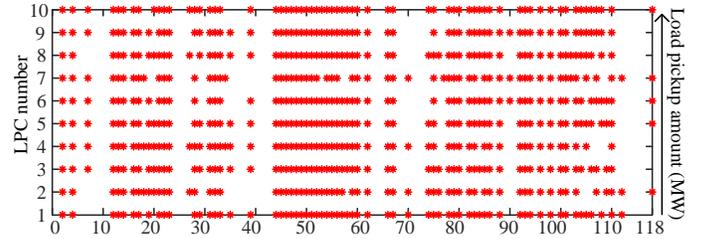

Fig. 8. Load checkup list for IEEE 118-bus system

*2) OSG based strategy generation*

TABLE III. LOAD RESTORATION PERFORMANCE OF IEEE 30-BUS SYSTEM

|  | Method | Step1 | Step2 | Step3 |
|---|---|---|---|---|
| Load pickup | OSG | 65.68% | 84.15% | 100% |
|  | C&CG | 63.18% | 83.81% | 100% |
| Comp. time | OSG | 0.08s | 0.11s | 0.09s |
|  | C&CG | 87.67s | 71.42s | 69.54s |

Applying the OSG algorithm, the well-trained DNN and CNN as well as the LPC will be used to realize the on-line robust load restoration strategy acquisition. The load restoration performance of the proposed model-free method is compared with the conventional C&CG based two-stage robust optimization in [12], [13]. Table III shows the multi-step performance of the IEEE 30-bus system. As can be seen, the restorations using both methods are all completed with three steps, while the model-free method is with slightly higher recovered load amount. Moreover, the model-free one has the obvious advantage in terms of calculation time. The model-based optimization method needs an average 76.21s to solve the two-stage robust load restoration model using C&CG algorithm, while the model-free one uses DL and LPC to achieve hundreds of times acceleration in computational time (Comp. time).

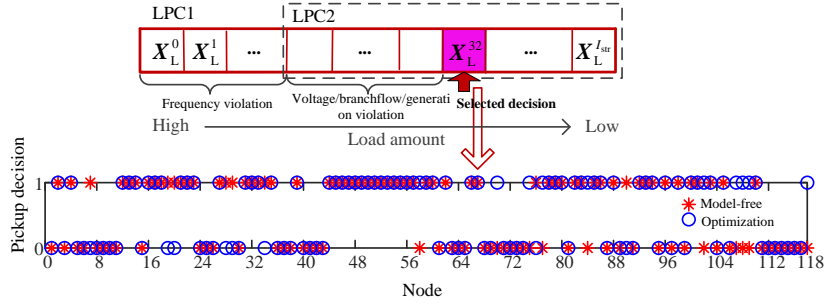

Fig. 9. Load pickup decisions of IEEE 118-bus system

The calculation benefit is more obvious with large-scale systems. Fig. 9 shows the load pickup decisions of model-based and model-free methods and the process of final decision acquisition of the IEEE 118-bus system. The final strategy is the 32nd set in the LPC, which is the result of deleting former sets because of frequency, voltage, branch flows and generation violations. The optimization method takes 1549.71 $s$ to reach the optimal robust load pickup strategy which restores 64 load blocks totaling 3577 MW. With the same security requirements, the proposed model-free method finds that the robust strategy restores 62 load blocks totaling 3588 MW in 0.24s. Compared with the optimization method, the model-free method adjusts load pickup decisions in 8 nodes with a 0.31% load amount increment. More important, it has more than 6000 times computational acceleration (Comp. accel) than the optimization one. Accordingly, the proposed model-free method has predominant computation speed and good optimization ability, which satisfies the on-line application.

TABLE IV. LOAD RESTORATION PERFORMANCE OF IEEE 118-BUS SYSTEM

|      | Load recovery | Recovered load blocks | Comp. time | Comp. accel. |
|------|---------------|----------------------|------------|--------------|
| OSG  | 3588MW        | 62                   | 0.24s      | 6457.13 times |
| C&CG | 3577MW        | 64                   | 1549.71s   |              |

## VI. CONCLUSION

The time-consuming calculation of the conventional model-based method limits the robust scheme to the off-line application. In order to realize on-line robust load restoration, this paper proposes a new DL based model-free method. A series of techniques, the DNN, deep CNN, LPCG algorithm and OSG algorithm, are developed and combined in order to completely replace the two-stage robust optimization. The designed DL method is with considerable accuracy for the worst-case condition capture and security check. The proposed LPCG algorithm helps maintain the optimization performance, and the OSG algorithm realizes fast on-line optimal robust load restoration acquisition. The case study results show distinguished computation benefits and satisfactory load restoration performance of the proposed method. The DL based model-free method fills the gap between the robust load restoration scheme and the on-line application, which further enhances bulk system resilience.